# Enhancing the Quality and Reliability of Machine Learning Interatomic Potentials through Better Reporting Practices

*Tristan Maxson, Ademola Soyemi, Benjamin W. J. Chen, Tibor Szilvási*[*]

[*]*Corresponding author. Email: tibor.szilvasi@ua.edu*

**Abstract**

Recent developments in machine learning interatomic potentials (MLIPs) have empowered even non-experts in machine learning to train MLIPs for accelerating materials simulations. However, the current literature lacks clear standards for documenting the use of MLIPs, which hinders the reproducibility and independent evaluation of the presented results. In this perspective, we aim to provide guidance on best practices for documenting MLIP use while walking the reader through the development and deployment of MLIPs including hardware and software requirements, generating training data, training models, validating predictions, and MLIP inference. We also suggest useful plotting practices and analyses to validate and boost confidence in the deployed models. Finally, we provide a step-by-step checklist for practitioners to use directly before publication to standardize the information to be reported. Overall, we hope that our work will encourage reliable and reproducible use of these MLIPs, which will accelerate their ability to make a positive impact in various disciplines including materials science, chemistry, and biology, among others.

## 1. Introduction

Computational analysis is an indispensable tool for understanding the behavior of materials. Such analysis has led to crucial insight in various fields such as catalysis[1-3], bio-[4-6], nano-[6-8], and soft[9, 10] materials among others, even enabling the computation-led design of materials in some cases. First-principles methods based on density functional theory (DFT) have been adopted by most communities to gain insights into materials properties due to a good balance between accuracy and computational cost. However, due to computational constraints, DFT is typically limited to simulating systems up to hundreds of atoms[11, 12] and time scales of up to hundreds of picoseconds[13, 14].

Recently, machine learning interatomic potentials (MLIPs) that are trained on DFT data have emerged as a promising method to expand the time and length scale of atomistic simulations while potentially maintaining DFT-like accuracy[15, 16]. Many ready-to-use software packages[17-28] are now available for prospective practitioners to download, install, and use within hours. The relative ease of using such packages has generated considerable interest, resulting in a rapid increase in scientific applications such as accelerating molecular dynamics[22, 29, 30], probing chemical reactivity[31-33], and investigating long-time- and length-scale phase transitions[34-36].

Due to this rapid growth in the field of MLIPs, standards for documenting the use of MLIPs have yet to be developed by the community, leading to inconsistencies in reporting on the developed MLIPs. Specifically, the materials modeling community has not reached a consensus on how to ensure reproducible accurate[37] use of MLIPs similar to what has been achieved for DFT-based simulations[38, 39]. As a result, widely different standards of reporting can be found in the literature ranging from merely mentioning the use of MLIPs to open-sourcing all data and scripts[40]. As MLIPs are still new and evolving, such differences are understandable. Nevertheless, their reliability can vary widely depending on MLIP construction and training procedures. Reproducing results from literature may be difficult even for experienced researchers. For example, simple physical interactions such as nuclear repulsion are not always included in models and atoms may collapse together unphysically. Improper training procedures may also overfit certain properties such as energy at the cost of forces, which may lead to unphysical dynamics. Thus, considering that current MLIPs are typically not robust constructions, there is a thin line between useful and misleading results. There is an urgent need for community to document the usage and validation of MLIPs with greater care to avoid misleading studies and over-interpreting results when there might be considerable uncertainty.

In this perspective, we aim to provide the community with guidelines for documenting the development, validation, and application of MLIPs (Figure 1). Specifically, we highlight different aspects of MLIPs that should be reported: hardware and software requirements, generation of training data, training of models, validation of predictions, and inference of MLIPs. In addition, we provide a detailed checklist that MLIP users can apply to self-review manuscripts before submission. Proper documentation of such information will foster greater trust in computational analyses and increase reproducibility. Promoting direct data availability will also accelerate the standardized development of MLIPs. Ultimately, developing better standards for detailing MLIP usage will help increase the adoption of MLIPs and allow the community to reap its full benefits.

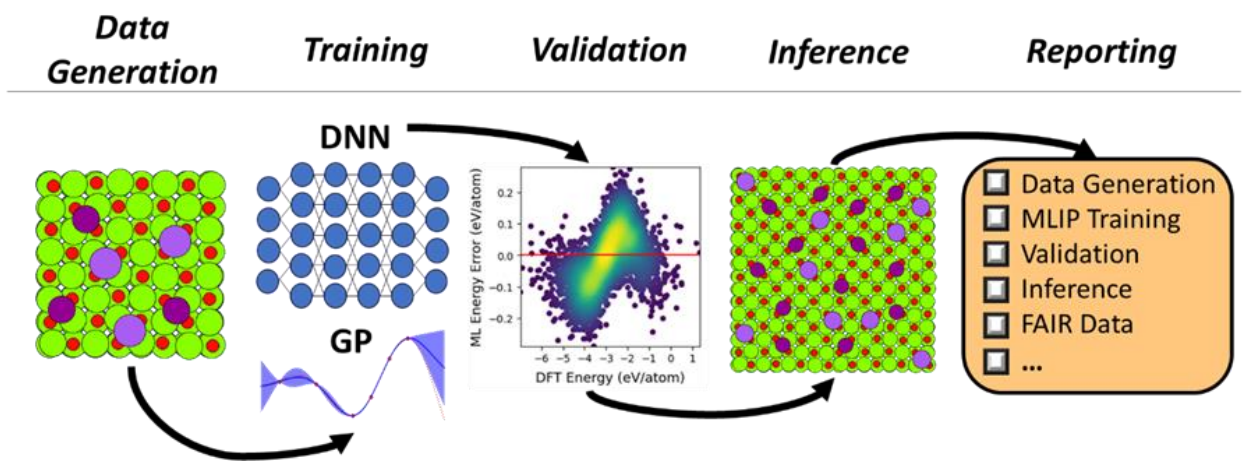

**Figure 1.** General workflow of data generation, training, validating, and deploying MLIPs for use in scientific research. Reporting is included as a final stage which is often overlooked in works utilizing MLIPs and is the focus of this perspective.

## 2. Best Practices for Reporting

### 2.1. Reporting Checklist

Based on the guidelines that will be discussed in this perspective, we have provided a convenient and comprehensive checklist of the process of documenting various aspects of MLIP usage, including the choice of software and hardware, training data generation, training, validation, and inference as seen in Figure 1. This checklist is provided in the Supporting Information (SI) for practitioners to go through before publication to help ensure reproducible research. The checklist also guides reviewers on what types of data should be present in submitted manuscripts, or at minimum aids in looking out for potentially missing information. An up-to-date version of the checklist may be found in the GitLab repository at https://gitlab.com/szilvasi-group/mlip-reporting-checklist. We encourage the community to make suggestions in the future as MLIPs continue to develop. The checklist may also be found in different formats if groups wish to improve the general checklist. We encourage MLIP developers to make pull requests for code-specific checklists if they wish to establish a more detailed standard for their own codes.

### 2.2. Software and Hardware

A range of software and hardware may be utilized by researchers. For software, we suggest reporting all aspects of the environment in which MLIPs are trained and used, such as the versions of the MLIP software package, Python, PyTorch/TensorFlow, CUDA, and the specifications of the used CPU/GPU. If an environment manager such as Conda or Python's venv is used, a versioned list of all packages installed may help track down potentially conflicting packages and allow others to reproduce the same environment.

It is also useful to note the type of hardware used as GPUs tend to be limited by memory requirements imposed by the size of the model. For example, an H100 GPU may have 80 or 188 GB while an A100 GPU may have 40 or 80 GB of memory. We suggest reporting detailed information regarding the GPU, CPU, memory, and parallelization to ensure that readers understand the viability of running the model.

**2.3. Training Data**

Training data for MLIPs may be generated and/or selected by a diverse range of methods. In this section, we detail various data generation strategies and their critical parameters to report. Training data typically consists of atomic configurations calculated with DFT, although any ground truth may be considered as well. The used atomistic data should be reported by providing the data source, version number, and access date that may be obtained from public repositories designed for MLIP development such as RevMD17[41], IonSolvR[41], and the Open Catalyst Project[42, 43]. Such repositories provide either the raw calculation files or the configurations in a compact format (xyz, npy, traj) including atomic positions, energy, forces, and stresses. If only a portion of the configurations were used, the subset of data used should be stated.

Existing molecular dynamics (MD) runs from the literature that were not originally intended for machine learning can also be possible sources of training data. The reuse of literature data can help may accelerate research projects in their initial stages, highlighting how the increased data availability may have positive effects beyond reproducibility. However, when no appropriate dataset is available for the specific use case, researchers are required to generate their training data since current MLIPs cannot typically generalize from one system to another unless there is a match in composition and conditions such as temperature and pressure. If this is so, the exact methods for generating new datasets should be rigorously documented.

This is especially so as new methods for data generation can advance the field and their value should not be underestimated by the researchers generating the training data and MLIPs[44]. While it is understandable that groups may have in-house tools that are not suitable for public sharing, the general methodology must be reported and is not restricted to the suggestions that follow in this section. Instead, we provide guidance on common methods and highlight critical parameters needed to reproduce the method. It is important to note that methods of generating training data should be viewed with equal value and disclosed as their reproducibility is not guaranteed by reporting the raw data.

When generating training data, it is helpful to start with simple and efficient methods, which usually involve enumerating different configurations of the system and evaluating them with single-point calculations. Yet, such methods can be surprisingly difficult to reproduce as critical details are often omitted. Three common methods for generating suitable atomic configurations involve random displacements[45, 46], phonon-like displacements[47-49], or combinatorial methods[50-55]. Random displacements, or structure rattling, perturb atomic positions to provide higher energy/force/stress structures from an optimized structure. Displaced structures provide a definition to the local PES but must be performed carefully and filtered to avoid generating unphysical configurations. Phonon displacements are similar to random displacements but are more physically motivated as they represent movements that may occur in MD simulations without directly performing them. Rattling structures or using phonons tends to be effective at exploring the local PES and it is helpful to provide the original structures before the displacement[47, 56-58]. For these displacement techniques, the type and magnitude of displacements, starting structures, unit cell, and phonon modes, should be provided.

Combinatorial methods involve the use of an algorithm to generate structures that ideally cover the entire configuration space required. The exact algorithm and method should be reported to identify possible biases, and a representative script should be provided. For example, unoptimized structures of liquids may be generated using a code such as Packmol[59] and thus an input file for Packmol[59] would be ideal to report. In contrast, crystalline materials such as high entropy alloys would require an entirely different algorithm with knowledge of a lattice structure to randomly place elements and distort the cell to reasonable lattice constants.[60] The SI should contain an example script or pseudocode that illustrates the logical flow and allows the reader to implement it in their choice of programming language. If the cost of generating additional structures is low, there may also be value in generating a supplementary dataset without calculated properties that can be used in the future to augment the main dataset.

In contrast to single-point methods, dynamical simulations can explore the potential energy surface directly and therefore give physically meaningful configurations for a given temperature, pressure, and composition. Reporting the MD ensemble (NVE, NVT, NPT), timestep, and thermostat/barostat settings is critical as these MD parameters directly affect the system's dynamics. Directly providing the input files required for the simulation is a bonus as it resolves any doubts about what has been done. Additionally, MD-based methods such as simulated annealing[61, 62], contour exploration[63], or minima/basin hopping[64-69] may be effective for exploring a wide variety of configurations. If such methods are used, their parameters must be reported consistent with literature studies utilizing them for molecular dynamics.

It may also be helpful to point out when simulations can be performed in parallel and when equilibration is not reached. When training MLIPs, it may not be necessary to perform MD runs that are suitable for analysis since the goal of the MLIP is typically to avoid performing long MD runs. Non-equilibrium MD should thus not be an issue when used as training data, especially as it is likely to improve the stability and robustness of the MLIP potential (stability itself is discussed in the Inference section) and biasing potentials that move the system away from equilibrium are helpful to use and report. If the bias is not directly implemented in the code being used, it is best to provide a small example of how to perform the biased MD[70].

Constraints in dynamical simulations can influence training data by intentionally biasing the system towards a given state and as such must be reported. Simple constraints such as the relationship between cell vectors or fixing atoms can be reported in the text and may be obvious in the trajectories supplied. However, it is crucial to provide unconstrained training data with all atoms having original forces restored, as calculated by the method used, even though some packages, such as ASE, zero the forces of fixed atoms by default. This is because some MLIP codes are not designed to read atomic constraints and the constrained atoms are included in the loss function with the constrained forces rather than the true forces. While some codes can mask atoms in the loss function to account for constraints, it is easier to simply store and use unconstrained data, which also ensures that the total energy and the forces in the system are fully consistent. More sophisticated constraint schemes (e.g., enhanced sampling[71-73]) must be reported as well and it is up to the author to properly describe them and/or provide an example script in the SI. For example, collective variables may be defined that control a bond distance as a function of some other property of the system to allow for a rare event to be performed that would be unlikely to occur naturally in an MD simulation.[74-77]

After generating the initial training data, it is also important to consider how relevant training points are selected. For example, using all frames of an MD simulation will result in highly correlated data. As such it is necessary to have a procedure for data selection, such as including only every $X^{th}$ atomic configuration. Alternatively, sampling configurations that cover a range of MD properties such as energy, force, pressure, or volume may avoid correlated structures.[70, 78] While more data is assumed to be helpful in improving an MLIP, our experience with modern codes suggests that smaller curated datasets tend to require fewer resources to train to high accuracy. If active learning is used, it will be necessary to note down the criteria for selection of sampled configurations[79-81]. Reporting the selection method and the unfiltered dataset allows for other selection methods to be tried later as they are discovered.

**2.4 Training MLIPs**

As model parameters may vary from package to package, detailed documentation of the model architecture, hyperparameters, and training methods is especially important for reproducibility. Although modern MLIP parameters might be confusing to those unfamiliar with specific packages, they can still be documented in a way that is generally useful and does not necessitate expert knowledge. For example, reporting of radial cutoff(s) allows for fundamental limits of the MLIP's knowledge of long-range order to be evaluated in a way that is completely agnostic to the code being used. We note, however, that while the radial cutoff determines the direct range of the MLIP, message-passing or non-local MLIPs may include contributions beyond this radial cutoff [82-84]. Similarly, a model's complexity can be estimated by its degrees of freedom with the caveat that not all degrees of freedom are equal within an MLIP or between different MLIPs. Most other MLIP parameters, however, may be crucial but difficult for non-expert readers to understand. We suggest including only the crucial parameters in the main document, especially those that have been explicitly tuned, while tabulating the rest in the SI. Another simple and efficient method for reporting is to disclose the full input file. We encourage MLIP developers to output files during training that can be directly used as input files. Such files should explicitly define all defaults to enable easy reporting in a way that is more agnostic to the specific version being used or if defaults are changed on installed package.

If changes are made to the publicly available version of the code and licensing permits, patches that are critical to training, together with their purpose, should be provided in the SI. Software improvements and implementation of alternative approaches should be viewed positively when they improve training; their value should not be underestimated even if they are simple.

While software packages may differ greatly, generic rules can be constructed for common models such as neural network (NN) potentials[18-21, 23-25, 27, 28]. The architecture and descriptors of a NN potential must be fully described with its layers, widths, and model features. The architecture of the model limits its accuracy given a fixed set of descriptors and as such determines the final quality of the model. Learning rates and loss functions may also influence the model strongly and any changes during training (multi-phase training, learning rate schedulers, etc.) should be detailed. Recently, the introduction of equivariance to MLIPs has also been seen to improve models[19, 85, 86]. For such codes, one should report related parameters such as the symmetries allowed or the maximum tensor ranks. Other general ML parameters such as activation function may not appear to be important but should be reported regardless as they may have unintended impact. If GPUs are used, it is important to report what numerical precision is used within the model as the usage of FP32, FP64, and TF32[87] may influence results depending on system size and normalization of data. If CPUs are used, FP64 should always be used (or higher precision) as most modern CPUs do not perform better on FP32 and precision will be lost unnecessarily. Precision in the model is unlikely to matter except when sub-meV/atom errors are reached[88]. The precision dependence can be evaluated by retraining or by reducing the model from FP64 to FP32 or TF32 in the inference stage.

Gaussian process potentials have also become common in the field especially for on-the-fly training [22, 89, 90] due to their predictable deterministic training time and inherent uncertainty quantification[91]. The greatest deviation from NN potentials is that they may contain differing radial cutoff definitions between atoms to represent 2-body, 3-body, or $n$-body terms and may not be as straightforward to add complexity to compared to an NN architecture. The radial cutoffs should be reported as a table such that particular interactions (e.g., C-H pairs) can be easily identified and compared with radial distribution functions. If calculated as part of the training process, the method of optimizing Gaussian process hyperparameters as well as the optimized hyperparameters for the model kernels and noise should also be reported to provide a good starting point for training similar models.

Gaussian processes are also convenient as the model may be precalculated in terms of its decomposed n-body interactions. The model is approximated by mapping the $n$-body terms directly to a tabulated form to be looked up during simulation, known as a mapped Gaussian process (MGP).[92] MGPs provide much faster predictions but may be less accurate if not determined on a fine enough grid. It is therefore critical to report the exact form of the approximation (e.g., grid spacing between tabulated points, linear vs. spline interpolation, error prediction, etc.). Plotting the 2 or 3 body terms may be helpful to identify issues at extreme distances. We also note that training data for Gaussian processes come from typical DFT calculations but are added on a per-atom basis rather than the entire configuration. The per-atom DFT data must be accounted for in the reporting. Training data may be better provided in the form of a list of configurations and atoms or by modifying the training dataset to include tags for atoms that are trained on. A standardized format for sparse atomic data which is supported across codes would be a useful development for the reproduction and cross-validation of codes. Any format can be used that supports tagging of atoms (extxyz, traj, npz) with an extra parameter that indicates to train on those atoms.

Hyperparameter optimization is commonly performed to search for the best models.[93] In such cases, it would be useful to note down the search space and the tabulate the tested parameters, which may save time and effort for prospective users looking to study similar systems. If online tools such as Weights & Biases are used for hyperparameter scans, the tools should also be mentioned together with the specific algorithms applied for hyperparameter optimizations[93-95].

## 2.5. Validating MLIPs

Evaluating the accuracy of an MLIP is commonly performed by assessing errors in uncorrelated structure predictions and in statistical/thermodynamic properties relevant to physical insights that will be drawn from the MLIP. The MLIP should predict energies, forces, and stresses accurately or any predictions will be a result of coincidence or error cancellation which may harm transferability or robustness[37, 96]. For proper validation, the test set must include carefully sampled data that is representative of the configurations the MLIP will need to predict although computational limitations may limit the system size. If that is the case, authors should report known limitations to inform the readers. A common error in creating test sets is to include atomic configurations from intermediate MD steps between those used as training images, which will give artificially low errors. Instead, we recommend building test sets from MD runs independent of those used for training so that the model's performance is evaluated on data that is not directly correlated with the training set. In addition, validation must not be performed using the "validation errors" as reported by MLIP codes as these errors are simply a metric to select an ideal model. The validation set leaks knowledge of its data into the model indirectly and influences the training process.

Error metrics, such as the mean absolute error (MAE), mean signed error (MSE), mean percentage error (MPE), or root mean square errors (RMSE), allow for easy quantification of the quality of the MLIP model and can highlight where systematic errors exist for improvement. Errors must be reported for all predicted properties such as energy, force, and stress. Force error can also be reported in terms of angle and magnitude, which are more meaningful and are rotationally invariant quantities. Force errors should also be computed based on the unmodified ground truth without constraints such as fixed atoms. Contributions from fixed atoms may be removed on a per-image basis instead but removing such contributions is not directly supported by all codes. Stress errors must also be reported carefully with the ideal gas contribution of velocities to stress removed before analysis as ideal gas contributions come from velocity in the MD simulation. The stress error should also be partitioned into the normal and shear components separately before analysis as it is common for shear stresses to not be considered during NPT simulations and the normal stress solely controls the pressure dependence.

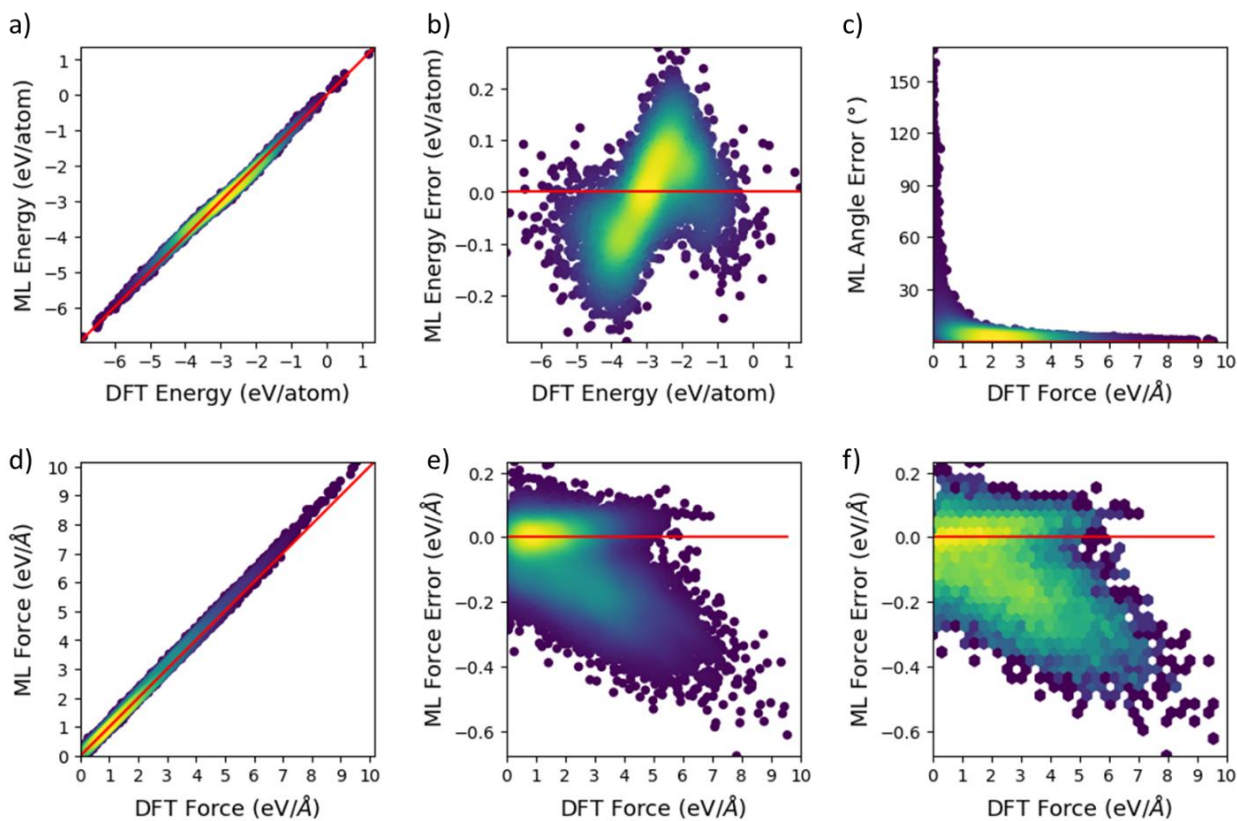

**Figure 2.** Improving the presentation of energy errors and force magnitude errors. Plots (a) and (d) are intended as examples of parity plots that show very little useful information, while plots (b) and (e) are the corresponding error distribution plots generated with KDE coloring highlight residual errors more effectively. The angle error of force is presented in plot (c). Plot (f) is the histogram analog of plot (e), showing the KDE plots are superior to a histogram due to distortion of data in the histogram. Data is generated as mock data from NumPy to highlight how to best plot MLIP errors and is not intended to be interpreted as a result of any simulation or training. A script to generate a similar plot is provided in the Supplementary Information. All plots are colored according to the default matplotlib color mapping (viridis).

Validation of MLIP potentials is often presented graphically as a function of system properties. For simple statistics such as energy or force errors, parity plots (Figures 2a and 2d) are typically the go-to choice as they allow the reader to easily compare predictions with benchmarks and determine if systematic errors are present. However, when parity plots are calculated over a wide range of values, the resolution of points is reduced, which may hide important discrepancies. In such cases, plotting the error distribution can help highlight specific patterns that cannot be seen on parity plots and indicate issues of the model (Figures 2b and 2e). For example, in Figure 2b there is a systematic deviation that is hidden in the parity plot of the same data. In Figure 2e, two distributions appear to be present, which may be due to atoms in different chemical environments or accidental mixing of training sets with different ground truths. Errors may also be correlated to other properties to highlight how predictions change as a function of the ground truth or predicted properties. For example, Figure 2c demonstrates a clear relationship between the force angle error and the magnitude of the force, with low magnitudes having larger angle errors in general. We suggest also plotting errors against properties such as pressure, atomic composition, and system size to check for correlations that might indicate simulation regions of concern and possible ways to improve the model.

To improve visual clarity when there is significant data overlap, the distribution of points can be colored via a kernel density estimation (KDE). This is visually clearer in comparison to a histogram, which more explicitly puts the data into bins to calculate a distribution but distorts the appearance as seen in Figure 2f. Note that the choice of bandwidth and kernel function can greatly affect the KDE, and as such these parameters should be reported as well. However, a histogram may be helpful if the data needs to be digitalized as discrete points may not be present when using a KDE in a region of high density. We also stress the importance of choosing a good color scheme when using color to encode data and suggest providing color bars and/or mentioning the color map if it has a common name. We provide an example script capable of producing both parity and error distribution plots in the SI.

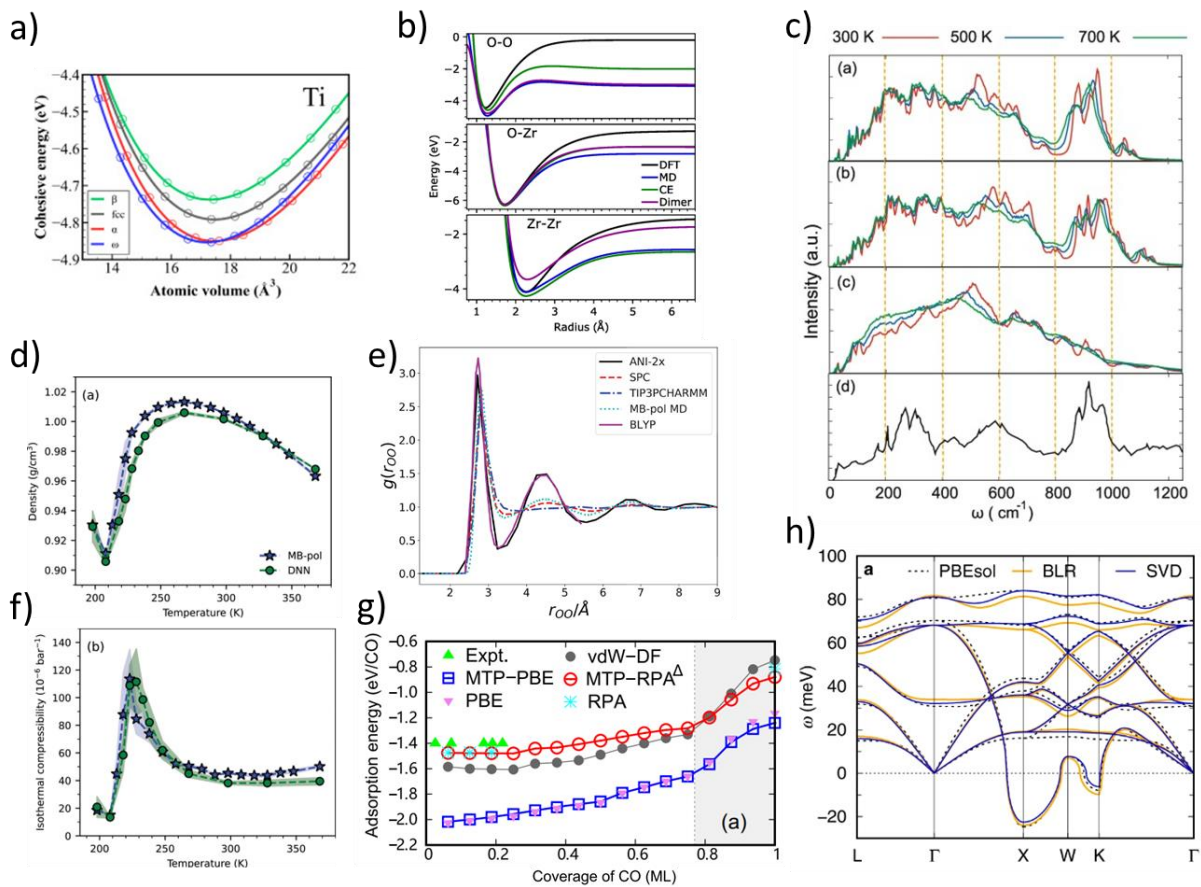

**Figure 3.** Example plots of (a) equation of state,[97] (b) two-body potential energy curves,[98] (c) vibrational density of states,[99] (d) density,[100] (e) radial distribution function,[101] (f) isothermal compressibility,[100] (g) coverage dependent adsorption energy,[102] and (h) phonon band structure[103] using MLIPs in literature studies. These highlighted plots demonstrate a wide variety of styles found in the literature for reporting MLIP predicted properties; we suggest the reader view them in the context of the original publication. (a) Adapted with permission[97]. Copyright 2022 Elsevier. (b) Adapted with permission[98]. (c) Adapted with permission[99]. Copyright 2022 Elsevier. (d, f) Adapted with permission[100]. Copyright 2023 AIP Publishing. (e) Adapted with permission[101]. Copyright 2021 American Chemical Society. (g) Adapted with permission[102]. Copyright 2023 American Physical Society. (h) Adapted with permission[103]. Copyright 2021 Springer Nature.

A particularly meaningful way to validate MLIPs is to check if the predictions in MD simulations match with the ground truth they are based upon. Such practice is especially useful when the validated property is directly related to the physical insights the MLIP study is intended to provide. Examples of MD-derived properties are the vibrational density of states, density, radial distribution function (RDF), and isothermal compressibility as seen in Figures 3c, 3d, 3e, and 3f, respectively. Evaluation of MD properties may be limited by equilibration time in DFT, so it may not be possible to directly validate with MD properties. Instead, it may be possible to equilibrate using the MLIP and report if a DFT simulation started from the final MLIP configuration starts to deviate in any thermodynamic properties over a short time, which may indicate a disagreement with the MLIP prediction. Tabulating all calculated properties in the SI will also facilitate direct comparison in the future when new methods or more comprehensive DFT simulations become available. We also suggest presenting differences of properties when the MLIP and DFT results appear to be well converged, since the resolution of data in plots may make it difficult to see where agreement is better or worse. For example, in Figure 3e the radial distribution function (RDF) appears to be very similar for the MLIP and MD runs. However, the agreement is not perfect, and a difference plot of RDFs would be informative. We note that the MLIP results should not be compared with experiments: this is a common mistake during validation of the MLIP and is addressed further in the inference section of the perspective.

Relevant thermodynamic quantities related to the studied research questions, such as *n*-body interactions, thermal expansion coefficients, binding energies, and phonons/vibrations, should also be evaluated and reported from analyses such as many-body decomposition[100, 104, 105] or analyses of static systems[106, 107] to ensure the MLIP provides meaningful physical insights. In many-body decomposition, DFT is used to calculate various configurations of 2-body, 3-body, or *n*-body isomers, enabling the precise determination of their decomposed interaction energies. Differences between MLIP and DFT interaction energies reveal the presence of systematic errors or instances of unphysical error cancellation. Two-body interactions may also be presented as a dissociation curve as seen in Figure 3b. Optimizing structures using DFT and MLIPs and reporting the residual mean square displacements (RMSD) can provide additional information about the low-energy structures predicted by the MLIP. Properties based solely on system energy, such as the formation energy of nanoparticles, cohesive energies of bulk structures, and binding energies of adsorbates, can be validated without running a full MD simulation at the DFT level. Other examples are an equation of state[108], the binding energy of adsorbates on a surface,[109] or the phonon band structure of a system[49, 110] as seen in Figures 3a, 3g, and 3h, respectively. We also suggest reporting validation of thermodynamic properties both statistically (for quick comparison) and graphically (for identification of trends) when possible.

Modern MLIPs are commonly strictly local potentials with their interactions confined to their radius of interaction, which guarantees that the models are size-consistent and size-extensive. However, in large unit cells, the long-range order that forms may be longer than the MLIP radius of interaction. On the other hand, in small unit cells, there can be artificial short-range order. The orders that form can thus bias the training set towards interactions that arise in specific system sizes and the MLIP may not generalize well as a result. It is logical to validate the system as a function of system size as well when the system size is changing. However, validating the model on large systems may present unique challenges, as MLIPs are typically used when the ground truth for larger systems is too expensive to compute. Therefore, when applying MLIPs trained on small systems to larger ones, it's advisable to invest in a limited set of single-point calculations to assess the MLIP's transferability to these larger systems. Potential errors may be verified and reported by checking for discontinuous or unphysical dissociation curves of atoms or by evaluating MD properties as a function of the number of atoms.

We also suggest reporting the test results of MLIPs beyond their training set conditions as extreme conditions, such as elevated temperatures, can help identify problems with stability and help us understand how the potential will extrapolate to unknown configurations[111, 112]. This test is particularly useful since the test accuracy of a model is not necessarily a good proxy of its extrapolative ability[111, 113]. Deliberately creating non-physical interactions in simulations may help identify when and why the MLIP fails during longer MD runs. For example, the O-H bond of water might break during MD runs, forming OH. If OH is not present in the training dataset, it would be good to include it to analyze how an unknown interaction will be handled in MD simulations if it occurs by chance. Similarly, performing a topology analysis in the water system would ensure that the simulation is still in a stable state given knowledge of the training dataset.

**2.6 Inferring MLIPs**

The most common use for MLIPs is to perform accelerated MD simulations in suitably modified classical MD software such as LAMMPS[114], GROMACS[115], or OpenMM[116]. Suggestions on how to perform and report MD simulations is beyond the scope of this perspective, as they should be performed identically to similar classical simulations. We do, however, suggest providing direct input files for reproduction. It can additionally be helpful for practitioners to know the MLIP evaluation speed, preferably reported in "atoms/second", as this is agnostic to system size and time step and may be related to the hardware reported. We also suggest reporting actual memory usage as the hardware provides a hard constraint on memory and requirements for inference may be significantly higher or lower than that for training depending on system size. Determining memory usage as a function of system size and providing this information in the SI will assist readers in understanding scaling of codes that they may be unfamiliar with.

MLIP inference is often challenging as major, hidden problems can arise due to stability issues[117]. While some guidelines for MLIP stability exist, addressing this complex issue is difficult as failures may occur differently on a code-to-code basis. MLIPs can sometimes predict unrealistic behavior even if the test dataset shows low errors as ML models typically do not follow physical constraints. For example, the MLIP might incorrectly predict bonding at unphysical distances or bonds may break unexpectedly. Such stability issues should be reported even if a solution may be out of reach. To improve stability, augmenting the MLIP with physically motivated terms such as nuclear repulsion, dispersive forces, and force constraints can help.[82, 118-120] If augmentations are applied, their exact form must be reported. It is also important to report any known limitations with regard to the temperature, pressure, or material phase they intended to operate in as it is common for stability issues to arise whenever the potential is extrapolating beyond its original training set.

Experimental comparison is often performed to validate trained MLIPs. This is a misguided practice as the underlying training data is not necessarily correlated with experiments: Errors arising from the computational method or model approximations will cause deviations from experiments. For example, we cannot consider a MLIP to be correct if it reproduces the experimental density of a material when the underlying DFT method does not. Therefore, MLIP results should not be directly compared and validated with experiments, but instead treated as emergent results from the underlying DFT method, given that the trained MLIP potential converges to the DFT results. As such, if the MLIP is expected to reproduce the DFT method well, any agreement with experiments likely indicates the underlying method would agree in absence of the use of MLIPs. Experimental comparisons are therefore useful for gaining physical insights into the underlying method, provided they are reported with caution to avoid grand claims that cannot be supported by the work.

## 2.7 Data Availability / FAIR

Data lies at the heart of machine learning. To promote software development and verify the results resulting from MLIPs, proper archival of data should be required. Data requests to authors may initially work but will be less effective as the publication and authors age. Adhering to the FAIR data principles[121-123]—making data findable, accessible, interoperable, and reusable—is most painless at the publication stage and must be advocated for by journals hosting MLIP publications. Public, unaltered datasets necessary for reproducing the results should be provided for publication. This should be provided in standard formats compatible with common software (Atomic Simulation Environment[124], pymatgen[125], etc.) to facilitate interoperability between codes. While the SI can serve as a backup of training inputs, large datasets are better hosted on research repositories like Zenodo[126] or NOMAD[127]. These repositories also provide DOIs for easy referencing of specific versions (which may be updated after publication) and retrieval of specific files from the dataset if properly formatted. We advise against using software-focused Git-based repositories such as GitHub and GitLab, since they may be modified less transparently, and bandwidth and size limitations may force authors to reduce the amount of data available.

Sharing trained models can also be very useful. As models are difficult to generally deploy across different computational systems without modification, we suggest providing all tools required for retraining. In some cases, the MLIP training code is not publicly available, making direct reproduction impossible. The models should still be reported as they will be useful for comparison with other MLIP training codes available at or after publication, encouraging healthy competition between development groups. Additional effort should be applied towards sharing trained models when training times are large due to the size of the dataset and retraining the model invokes an excessive cost on future practitioners preventing reproduction. An open interface online which is provided for a pre-trained model, such as that of the Open Catalyst project demo on the OC20 dataset[42, 128, 129], can be a helpful alternative when sharing the model is not possible. However, there must be still caution with the suggested practices as there is no way to guarantee that third-party services hosted by individual researchers will be available long term. If it is possible, providing the repository containing everything required to self-host the service helps ensure it is not lost if the original service provider is no longer functional.

## 3. Conclusions

Establishing clear and comprehensive reporting standards for MLIPs is critical for ensuring the reliability, reproducibility, and advancement of MLIP-accelerated simulations. This perspective highlights various aspects that may not currently be rigorously documented but are critical for increasing the usefulness and reproducibility of publications. We encourage the adoption of standardized reporting practices that are enforced at the personal, group, and journal levels before publication. A more standardized approach will also assist in peer review by ensuring that reviewers have all pertinent information on their first viewing, accelerating the publication process by reducing the need for lengthy revisions. The challenges with MLIP reporting arise largely from the diverse reporting requirements of newer software available and the lack of long-term, field-specific standards. To tackle this, we provide a checklist in the Supporting Information encompassing the suggestions of this perspective. The dynamic and evolving nature of MLIPs and their applications requires an adaptable standard that lays the groundwork for all MLIPs to be reported similarly. We hope that the establishment of such standards may also accelerate the adoption of MLIPs in the common workflows of DFT practitioners.

**Supporting Information**.

The following files are available free of charge.

Checklist of parameters to report for authors to reference prior to publication and a script to serve as an example of how to plot validation data using matplotlib. (PDF)

**Author Information**

**Corresponding Author**

**Tibor Szilvási** - Department of Chemical and Biological Engineering, University of Alabama, Tuscaloosa, AL 35487, United States; Email: tibor.szilvasi@ua.edu

**Authors**

**Tristan Maxson** - Department of Chemical and Biological Engineering, University of Alabama, Tuscaloosa, AL 35487, United States

**Ademola Soyemi** - Department of Chemical and Biological Engineering, University of Alabama, Tuscaloosa, AL 35487, United States

**Benjamin W. J. Chen** - Institute of High Performance Computing (IHPC), Agency for Science, Technology, and Research (A*STAR), 1 Fusionopolis Way, #16–16 Connexis, Singapore 138632, Singapore

**Author Contributions**

The manuscript was written through contributions of all authors. All authors have given approval to the final version of the manuscript. All authors have contributed equally.

**Notes**

The authors declare no competing financial interest.

**Acknowledgements**

T.M., A.S. and T.S. would like to acknowledge the financial support of the National Science Foundation (NSF) under grant number 2245120 and the financial support of the Department of Energy (DOE) under grant number DE-SC0024654. T.M. would like to acknowledge this material is based upon work supported by the U.S. Department of Energy, Office of Science, Office of Advanced Scientific Computing Research, Department of Energy Computational Science Graduate Fellowship under Award Number(s) DE-SC0023112. A.S would like to acknowledge the financial support of the University of Alabama Graduate School as a Graduate Council Fellow. B.W.J.C is grateful for support by the A*STAR SERC Central Research Fund award. T.M., A.S. and T.S. would also like to thank the University of Alabama and the Office of Information Technology for providing high-performance computing resources and support that has contributed to these research results. This work was also made possible in part by a grant of high-performance computing resources and technical support from the Alabama Supercomputer Authority. This research used resources of the National Energy Research Scientific Computing Center (NERSC), a U.S. Department of Energy Office of Science User Facility located at Lawrence Berkeley National Laboratory, operated under Contract No. DE-AC02-05CH11231 using NERSC award BES-ERCAP0024218. Any opinions, findings, conclusions, and/or recommendations expressed in this material are those of the authors(s) and do not necessarily reflect the views of the NSF or DOE.

**TOC:**

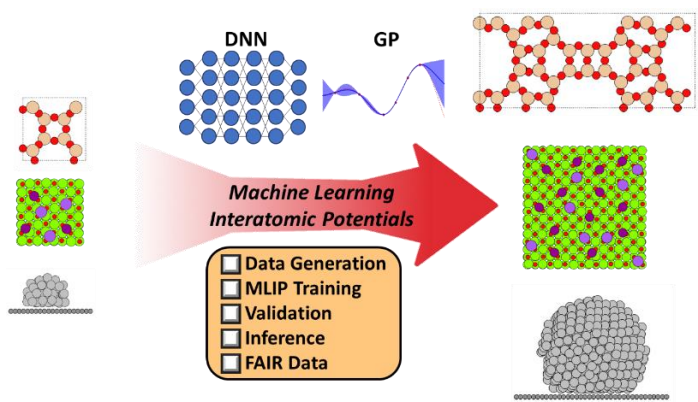